\documentclass[12pt,onecolumn]{IEEEtran}
\usepackage{amsmath,graphicx,amssymb, amsfonts, amsthm, color, subfigure}

\newtheorem{theorem}{Theorem}

\newtheorem{lemma}{Lemma}

\newcommand{\graph}{{\cal   G}}

\newcommand{\dummy}{N}

\newcommand{\Cp}{C}
\newcommand{\Rate}{R}

\newcommand{\maxcap}{c}

\newcommand{\dg}{d}

\newcommand{\block}{{n}}
\newcommand{\Block}{{\block}}
\newcommand{\Bl}{{\Block}}
\newcommand{\F}{{\mathbb {F}}}

\newcommand{\cantor}{{K}}

\newcommand{\tran}{{T}}

\newcommand{\eX}{{X}}
\newcommand{\eY}{{Y}}
\newcommand{\vX}{{\mathbf{X}}}
\newcommand{\vY}{{\mathbf{Y}}}

\newcommand{\Edges}{{\mathcal E}}
\newcommand{\NumE}{{|\Edges|}}

\newcommand{\dep}{{\Delta}}

\newcommand{\Vertices}{{\mathcal V}}
\newcommand{\NumV}{{|\Vertices|}}

\newcommand{\source}{s}

\newcommand{\edge}{e}
\newcommand{\nodeu}{u}
\newcommand{\nodev}{v}
\newcommand{\nodew}{w}

\newcommand{\sink}{t}
\newcommand{\sinks}{{\mathcal T}}
\newcommand{\NumS}{{|\sinks|}}

\newcommand{\wupp}{{\bf WUP($\epsilon,\NumS$) }}
\newcommand{\supp}{{\bf SUP($\epsilon$) }}
\newcommand{\rtdt}{{\bf R2-}${\mathbf D^2}$ }
\newcommand{\ctpo}{{\bf C3-}${\mathbf P0}$ }

\newcommand{\cP}{{\cal{P}}}

\newcommand{\flowcut}{F}
\newcommand{\flowset}{{\mathcal{F}}}

\begin{document}
\title{Universal and Robust Distributed Network Codes}

\author{
\authorblockN{Tracey Ho{\textdagger}, Sidharth Jaggi{\textdaggerdbl}, Svitlana Vyetrenko{\textdagger} and Lingxiao Xia{\textdaggerdbl}}
\authorblockA{{\textdagger} California Institute of Technology, {\textdaggerdbl} The Chinese University of Hong Kong\\
{{\textdagger}\{tho, svitlana\}@caltech.edu, {\textdaggerdbl}\{jaggi, lingxiao\}@ie.cuhk.edu.hk}
}
}
\maketitle
\begin{abstract}
Random linear network codes can be designed and implemented in a distributed manner, with low computational complexity. However, these codes are classically implemented~\cite{HoKMKE:03} over finite fields whose size depends on some global network parameters (size of the network, the number of sinks) that may not be known prior to code design. Also, if new nodes join the entire network code may have to be redesigned.

In this work, we present the {\it first universal and robust} distributed linear network coding schemes. Our schemes are universal since they are independent of all network parameters.
They are robust since if nodes join or leave, the remaining nodes do not need to change their coding operations and the receivers can still decode. They are distributed since nodes need only have topological information about the part of the network upstream of them, which can be naturally streamed as part of the communication protocol.

{
We present both probabilistic and deterministic schemes that are all asymptotically rate-optimal in the coding block-length, and have guarantees of correctness. Our probabilistic designs are computationally efficient, with order-optimal complexity. Our deterministic designs guarantee zero error decoding, albeit via codes with high computational complexity in general. 
}
Our coding schemes are based on network codes over ``scalable fields". Instead of choosing coding coefficients from one field at every node as in~\cite{HoKMKE:03}, each node uses linear coding operations over an ``effective field-size" that depends on the node's distance from the source node. The analysis of our schemes requires technical tools that may be of independent interest. In particular, we generalize the Schwartz-Zippel lemma~\cite{MotRag95} by proving a non-uniform version, wherein variables are chosen from sets of possibly different sizes. We also provide a novel robust distributed algorithm to assign unique IDs to network nodes.
%


\end{abstract}
\section{Introduction}
The paradigm of network coding allows each node in a network to process information in a non-trivial manner. As shown in~\cite{Ahlswede++2000,Li++2003, KoeM:02}, even if intermediate nodes simply perform linear operations over some finite field, the resulting {\it network codes} can be information-theoretically rate-optimal for a large class of communication problems. In particular algorithms that design codes for {\it multicast} communication problems, wherein each of multiple sinks requires the same information from a source node, have been well-studied. The design algorithms in~\cite{JagSanCEEJT, LangbergSB:05} are deterministic and centralized, and result in network codes with zero-error. In contrast the algorithms in~\cite{HoKMKE:03,JagCJ:03} are decentralized and probabilistic, and for any $\epsilon>0$ result in network codes that ``fail" with probability at most $\epsilon$. Both these types of design algorithms (and the resulting network codes) are computationally tractable.

However, for all current network code design algorithms, some information network parameters is necessary prior to the code design, to determine the size of the finite field over which linear network coding is performed. In particular, the centralized algorithms in~\cite{JagSanCEEJT, LangbergSB:05} require prior knowledge of the entire network, and even the decentralized algorithms in~\cite{HoKMKE:03,JagCJ:03} require knowledge of the network size and the number of sinks -- if these parameters are unavailable, code design cannot proceed with any guarantees of correctness, hence prior designs are not universal. Also, in the case of dynamically changing network topologies, if even one new node joins the network the entire network code may need to be updated due to a change in the field-size required, hence such codes are also not robust.

In this work we develop the {\it first universal and robust} distributed linear codes that are  independent of all network parameters, and are designed to satisfy a pre-specified tolerance on the error-probability (defined as the probability that the linear transform from the source to some sink is not invertible). 
The essential idea behind our design is that of using ``scalable fields"\footnote{Scalable fields do {\it not} mean embeddings of, for example, $\F_2$ into $\F_{2^2}$, since arithmetic operations are defined differently over different fields, and hence the overall transform would not be linear.
}. Linear coding operations are chosen from nested finite subsets of an appropriate infinite field -- in particular we choose $\F_2(z)$ the field of {\it rational functions over $\F_2$}, {\it i.e.}, the field whose elements are ratios of binary polynomials. Operations over this field can be implemented via binary filters (or equivalently, convolutional codes) at each node. For instance, a node that chooses to implment the operation $1+z^2$ on an incoming binary sequence $x(n)$ to generate an outgoing binary sequence $y(n)$ would set $y(n) = x(n) + x(n-2)$. Convolutional network coding as a model of linear network coding has been well-studied (see for example~\cite{ErezF04}).

As information percolates down the network, each node makes its own estimate of the ``effective field size", {\it i.e.}, the size of the subset of $\F_2(z)$ from which that node should choose its coding operations, so as to meet the guarantee on the pre-specified tolerance on the overall error-probability. Our codes are able to perform this book-keeping despite having access only to information that can be percolated down the network at rates that are asymptotically negligible in the block-length -- like standard distributed network codes, our codes are also asymptotically rate-optimal.

Our results are as follows. In Section~\ref{sec:Lemmas} we prove a generalization of the Schwartz-Zippel lemma~\cite{MotRag95} that is useful as a technical tool in some of our code constructions; it may also be of independent interest for other universal algorithms. 

In Section~\ref{sec:prob} we present probabilistic universal and robust codes. That is, given any $\epsilon > 0$ and any network, we present codes that guarantee that the linear transform from the source to each sink is invertible with probability at least $1-\epsilon$ -- hence our codes are universal. Further, even if nodes join or leave, pre-existing nodes do not need to change their coding operations to preserve the same guarantee of correctness -- hence our codes are robust. We present two such codes. The first code is independent of network size, but does depend on the number of sinks. We present it primarily for expository purposes, since its presentation is simpler than that of our second set of codes, which are independent of all network parameters, including the number of sinks. Both these sets of codes  base their choices of coding operations based on their distance from the source node.
While the effective field-size over which our codes operate, and hence the computational complexity of our codes, are larger than those of prior distributed designs~\cite{HoKMKE:03,JagCJ:03}, the complexity of implementing them is still polynomial in network parameters. Also, 
we present in Theorem~\ref{thm:lower_bound} a class of networks that demonstrates that our codes have essentially order-optimal computational complexity for universal codes.

In Section~\ref{sec:starwars} we consider deterministic universal and robust codes. As a technical tool we first discuss a decentralized algorithm to distribute unique IDs to each node in a robust manner -- even if a new node joins we guarantee that it too can be given an ID that is distinct from all others in the network. Building on this tool, and a novel use of Cantor's classical mapping between ${\mathbb Z}$ and ${\mathbb Z}^n$ for any finite $n$, we design zero-error decentralized codes that are independent of all network parameters, and robust to changes in network topology. We provide two constructions. Our first construction, also primarily expository, is just for codes of rate $2$, and is computationally efficient to design and implement. Our second construction is for arbitrary rate codes. This generalization comes at the cost of exponentially increasing the implementation complexity, compared to our other constructions\footnote{We distinguish between the computational complexity of design and that of implementation. The former refers to the the computational cost of designing the coding operations at each node, and is a one-time cost. The latter corresponds to the computational cost incurred by each node as it implements the pre-designed coding operations, and is a repeated cost for each packet transmitted by that node. All our codes have design complexity that is at most polynomial in the network parameters. Further, most of our designed codes codes have implementation complexity that is also polynomial in network parameters; the only exception is the last of the proposed designs corresponding to the general design for zero-error universal and robust distributed linear codes. 
}.

We note that all our algorithms provide guarantees of correctness as long as the source transmits information at a rate no greater than can be supported by the network, {\it i.e.}, its min-cut. We view the process of determining this rate as a rate-control issue -- our code designs are independent of the size of the min-cut.

\subsection{Related work}
The distributed random linear codes of~\cite{HoKMKE:03, JagCJ:03} require field-sizes to scale roughly as $\NumE\NumS$. As shown in~\cite{LL04}, even with centralized design of network codes, the field size over which coding must be performed as at least $\NumS$. 

As to universal codes (codes independent of some problem parameters), they have been well-studied in the classical information-theory setting (for instance in source coding~\cite{Rissanen:83} and channel coding~\cite{CsiK:81}).

{In the network coding setting, however, the literature is much sparser. The work of~\cite{KoeM:02} proposes ``robust network codes" that are resilient to network failure patterns. However, the field-size over which coding is performed depends on the number of failure patterns, and hence these codes are not truly universal. Further, the computational complexity of designing such codes is prohibitive. There is also significant work on network coding for packet erasure networks (for instance~\cite{DanGPHE:06}). 
Our codes can tolerate all such errors.

The work of~\cite{FraS:04} examines ``decentralized network coding" in which new nodes can join a network without disrupting pre-designed coding operations. Here, too, the field-size choice for the initial design depends on the size of the network. Further, the code designs are for special cases -- either for rate-$2$ codes (analogous to the codes we present in Section~\ref{subsec:r2d2}) or for networks with only two sink nodes.

A preliminary version of the work in Section~\ref{sec:starwars} was previously in the thesis~\cite{Jaggi:06}, and presented (but not published) in~\cite{JagHE:07}.
}

\section{Notation and definitions}
\subsection{Network Model}
In this paper, we adopt the single-source multicast network model of~\cite{KoeM:02}. 
Let the network be represented by a directed acyclic\footnote{The work can be directly extended to multi-source multicast networks, and over networks that may contain cycles, as long as each source has a unique identifier. To ease notational and description complexity we omit details here}
 graph $\graph=(\Vertices,\Edges)$. Here $\Vertices$ represents the set of nodes and $\Edges$ the set of edges. The graph has a pre-specified source node $\source$ and $\NumS$ sink nodes $\sinks=\{\sink_1, \sink_2, \ldots, \sink_{\NumS}\}$. 
A directed edge $e$ from node $\nodeu$ to node $\nodev$ is said to have {\it tail} $\nodeu$ (denoted $tail(\edge)$) and {\it head} $\nodev$ (denoted $head(\edge)$).
The link $\edge$ is then said to be an {\it incident outgoing link} of $\nodeu$ and an {\it incident incoming link} of $\nodev$.

\subsection{Communication Model}
The communication goal is for the source to communicate identical information to each sink.

As is standard~\cite{KoeM:02}, we assume that each link carries one packet of information per time-step. This is reasonable since if some link's capacity is less we may consider the link's communication to be over multiple successive time-steps, and if the link's capacity is greater we can subdivide it into multiple links. The {\it packet-length} in bits is denoted by $\Bl$.

The {\em network capacity}, denoted by $\Cp$, is the
time-average of the maximum number of packets that can be delivered from the source $\source$ to each sink $\sink \in \sinks$ simultaneously.
It can be also expressed as {\em the minimum of the min-cut from the source $\source$ to each sink $\sink$}.
The rate $\Rate$
is the average number of {\em information} packets that the source $\source$ generates per time-step, to be delivered to each sink $\sink$ over the network $\graph$. Without loss of generality we assume that $\Rate < \Cp$. {Lastly, let $\maxcap$ denote the maximum capacity of any single link in the network.}

\subsection{Code Model}

\subsubsection{Network code}
The {\it network code} comprises of the encoders at the source and each node inside the network, and the decoders at each of the sinks. In particular we focus on {\it linear network codes}, {\it i.e.}, codes where the source node, each internal node, and each sink performs linear combinations of information in packets on incident incoming links to generate packets on incident outgoing links. Specifically we consider the class of {\it convolutional} linear operations, well-studied in classical coding theory, that we reprise below. The base-field for arithmetic is chosen to be $\F_2$, hence all operations described below are binary.

\subsubsection{Convolutional network code}
Recall that the {\it $z$-transform}~\cite{MacKay:03} of any sequence $\{a(i) \} _{i=1}^\Bl$ of bits is given by the polynomial $\sum_{i=1}^\Bl a(i)z^i$, denoted $A(z)$. Further, recall that the output of the {\it convolution operation $a \ast b$} between two sequences\footnote{Terms $a(i)$ and $b(j)$ are respectively set to zero for $i \notin \{1,2,\ldots,\Bl\}$ and $j \notin \{1,2,\ldots,\Bl'\}$.} $\{a(i) \} _{i=1}^\Bl$ and $\{b(i) \} _{i=1}^{\Bl'}$ is defined as the length-$\Bl + \Bl'$ sequence whose $i$th term equals $\sum_{j=1}^{i}a(j)b(\Bl'-j+i)$. Lastly, it is well-known that the $z$-transform of $a \ast b$ equals $A(z)B(z)$. 

Convolutional codes~\cite{MacKay:03} have long been used in point-to-point communication scenarios. The idea of using convolutional codes for network coding (in networks with cycles) was foreshadowed in~\cite{Ahlswede++2000}, and made explicit in~\cite{KoeM:02} (who also noted that such an algebraic model for coding operations can help kill two birds with one stone, {\it i.e.}, it can also help model delays in networks). The work of Erez {\it et al.} (see for example~\cite{ErezF04}) gave the first efficient designs for convolutional network codes, {\it i.e.}, codes over $\F_2(z)$. In our work, $\F_2(z)$ affords the advantage that it allows for coding operations to be chosen from a potentially unbounded set, which helps us circumvent the difficulty that we do not know the network's parameters in advance.

The source's packets are denoted by $\eX_1,\eX_2,\ldots,\eX_\Rate$ -- each is a length-$\Bl$ bit-vector. The corresponding $z$-transforms are denoted $\eX_1(z),\eX_2(z),\ldots,\eX_\Rate(z)$. Collectively they are represented by the length-$\Rate$ vector of polynomials $\vX(z)$.
Each edge $e$ carries the packet $\eY_\edge$, and its $z$-transform is denoted $\eY_\edge(z)$. Lastly, the $z$-transform of the packets on incident incoming links to any sink $\sink$ are denoted by the vector $\vY_\sink(z)$. We henceforth refer to a sequence and its $z$-transform interchangeably. 

Let $\nodeu$, $\nodev$ and $\nodew$ be three nodes such that there is at least one edge from $\nodeu$ to $\nodev$ and at least one edge from $\nodev$ to $\nodew$. We use a $5$-tuple to denote a coding choice for such nodes -- specifically, $\beta_{\nodeu,i,\nodev,j,\nodew}(z)$ refers to the {\it local coding coefficient} of the convolutional operation on the information on the $i$th edge from $\nodeu$ to $\nodev$ to the $j$th edge from $\nodev$ to $\nodew$. The choices of values of the local coding coefficients $\beta_{\nodeu,i,\nodev,j,\nodew}(z)$ are code design parameters whose specifications are the primary objective of subsequent sections.
Let $\edge$ be a specific edge from $\nodev$ to $\nodew$, and $\edge'$ denote a dummy variable that ranges over all edges incoming to $\nodev$ (and hence is indexed by the pair $(\nodeu,i)$). Thus the convolutional operation that is performed at node $\nodev$ comprises of taking linear combinations of the information $\eY_{\edge'}$ with the appropriate $\beta_{\nodeu,i,\nodev,j,\nodew}(z)$ over all edges $\edge'$ incident incoming to node $\nodev$, to generate the information $\eY_{\edge}$ on the edge $\edge$ incident outgoing from $\nodev$. 
To simplify notation, we henceforth write $\beta_{\nodeu,i,\nodev,j,\nodew}(z)$ simply as $\beta_{\edge,\edge'}(z)$ with the understanding that $(\edge,\edge')$ index the appropriate $5$-tuple. Thus the linear transform at each node can be written symbolically as
$$ \eY_\edge(z) = \sum_{\edge'} \beta_{\edge',\edge}(z)\eY_{\edge'}(z).
$$
Since all the linear operations performed by the network can be represented via operations over polynomials over the binary field, we henceforth consider all arithmetic to be over the {\it field of rational functions}~\cite{KoeM:02} over $\F_2$, denoted by $\F_2(z)$. The elements of this field are of the form $P(z)/Q(z)$, where both $P(z)$ and $Q(z)$ are binary polynomials. Linear codes over this field have been well-studied in the convolutional coding literature~\cite{MacKay:03}.

As in classical distributed network codes~\cite{HoKMKE:03}, the codes in this work are {\it distributed}, {\it i.e.}, the choice of a value for $\beta_{\nodeu,i,\nodev,j,\nodew}(z)$ at node $\nodev$ can depend only on its local parameters $(\nodeu,i,\nodev,j,\nodew)$, and the corresponding parameters of the nodes upstream of node $\nodev$. Since we consider only directed acyclic networks in this work, this imposes a significant design constraint, since nodes that cannot directly communicate with each other over the network cannot coordinate their coding choices.

One idea of~\cite{HoKMKE:03} that we too use is the idea of having ``short headers" in each transmitted packet. Specifically, each packet (containing $\Bl$ bits) transmitted by the source, also contains the linear transformation induced by the network from the source to that packet -- as in~\cite{HoKMKE:03} these transforms are computed in a distributed manner and percolated down the network along with the payload information at an asymptotically negligible rate.
For every $\sink \in \sinks$, let $\tran_\sink$ be the {\it network transfer matrix} from $\source$ to $\sink$ -- these too can be computed in a distributed manner. Let $\tran$ be the {\it overall network transfer matrix} from $\source$ to $\sinks$ formed as $\Pi_{\sink \in \sinks}\tran_\sink$. 
Let  $|\tran_\sink|$ and $|\tran|$ denote the determinants of $\tran_\sink$ and $\tran$ respectively.



Our codes are either  {\it probabilistic} or {\it deterministic} depending on whether local coding coefficients are chosen probabilistically or deterministically.
The {\em error probability} is the probability over choices of local coding coefficients that for each source message, at least one sink's reconstruction of at least one possible message from the source is inaccurate. For linear network codes this happens if and only if the transfer matrix from the source to each sink is invertible.
Rate
$\Rate$ is said to be achievable if for any $\epsilon>0$ and
$\delta>0$ there exists a coding scheme of block length $\Block$
with rate $\geq \Rate-\delta$ and error probability $\leq \epsilon$.
In particular, we require our deterministic network codes to be {\it zero-error}, {\it i.e.}, to have error probability be zero.




\section{The Generalized Schwartz-Zippel Lemma}
\label{sec:Lemmas}

The classical Schwartz-Zippel lemma~\cite{MotRag95} provides an upper bound on the probability that when variables of a polynomial are chosen uniformly at random from a field, then the polynomial evaluates to zero. 

Recall that the {\it degree $\dg_i$ of a variable $x_i$ in a polynomial $P(x_1,x_2,\dots,x_\dummy)$} is the maximal exponents of $x_i$ in its non-zero terms. Further, recall that the {\it degree $\dg$ of a polynomial $P(x_1,x_2,\dots,x_\dummy)$} itself is the maximal value among the sum of the exponents of all its non-zero terms. Note that $\dg \leq \sum \dg_i$.

\begin{lemma} [Schwartz-Zippel lemma~\cite{MotRag95}]
\label{lem:SZ_Lemma}
Let $P(x_1,x_2,\dots,x_\dummy)$ 
be a non-zero polynomial of degree $d\geq 0$ over a field $\F$. Let $S$ be a finite subset of $\F$, and the value of each $x_1, x_2,\dots, x_\dummy$ be selected independently and uniformly at random from $S$. Then the probability that the polynomial equals zero is at most
${\dg}/{|S|}.$
\end{lemma}
\noindent The Schwartz-Zippel lemma is a useful tool in the analysis of random linear network codes (for instance~\cite{HoKMKE:03}).
A random linear network code causes an error if and only if one of the transfer matrices from the source to the destination is singular. This in turn happens if and only if the product of the determinants of these transfer matrices equals zero. But this product of determinants may be viewed as a polynomial whose variables consist of the local coding coefficients at each node. Hence the Schwartz-Zippel lemma provides an upper bound on the probability of error of a random linear network code. 

In this work we are interested in a generalization of the Schwartz-Zippel lemma, for polynomials whose variables are chosen from different subsets of $\F$. We prove:
\begin{lemma}
\label{lem:SZ_Extend1}
Let $P(x_1,x_2,\dots,x_\dummy)$ 
be a non-zero polynomial over a field $\F$. For all $i\in \{1,2,\ldots,\dummy\}$, let $S_i$ be a finite subset of $\F$, the degree of $x_i$ in $P(x_1,x_2,\dots,x_\dummy)$ be $\dg_i$, and the value of each
$x_i$ be selected independently and uniformly at random from $S_i$. Then the probability that the polynomial equals zero is at most
$\sum_{i=1}^\dummy \left ({\dg_i}/{|S_i|} \right ).$
%
%
%
\end{lemma}
\noindent {\bf Proof:}
Given in Appendix~\ref{sec:app1}
\hfill $\Box$\newline
\noindent {\bf Note $1$:} Neither Lemma~\ref{lem:SZ_Lemma} nor Lemma~\ref{lem:SZ_Extend1} put any restriction on the size of the field $\F$, as long as the appropriate subsets from which variables chosen are finite.\newline
{\noindent {\bf Note $2$:} A related but inequivalent generalization of the Schwartz-Zippel lemma was proved in~\cite{Alo:99a}.}
\newline
\noindent The utility of this proof is that it allows for the variables comprising the polynomial to be chosen non-uniformly. This is integral to the proof techniques in this work, wherein we choose local coding coefficients over progressively larger sets, depending on how far from the source they are.
%
%
{\section{Probabilistic designs}\label{sec:prob}
In this section we describe two probabilistic designs of universal distributed robust network codes. 
In particular, given any $\epsilon > 0$, we present schemes such that the overall error probability of the code is at most $\epsilon$.}

Our first scheme is independent of the size (number of nodes/edges) of the network, but does requires that the source has {\it a priori} knowledge of the number of sinks it shall be required to service. Hence we say it is only {\it weakly universal}. 
Our purpose in presenting this scheme is primarily expository, since the proof is significantly easier than that of the second scheme -- it helps set the stage for the second scheme.

The second scheme is {\it strongly universal} and is independent of all network parameters, including the number of sinks.

We first describe some useful preprocessing steps relevant for both of our schemes.
\subsection{Graph transformation}\label{subsec:transform}
\begin{figure}[h]
   \begin{center}
   \subfigure[Coding node.]
   {\scalebox{0.5}{\includegraphics{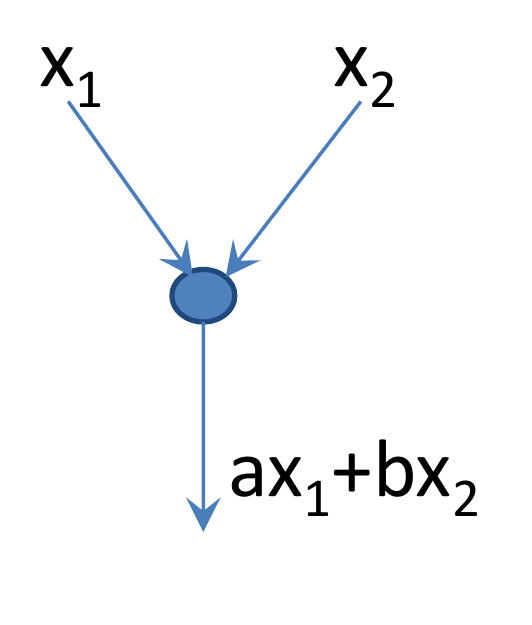}}
   \label{figure:coding}}
\subfigure[Broadcasting node.]
   {\scalebox{0.5}{\includegraphics{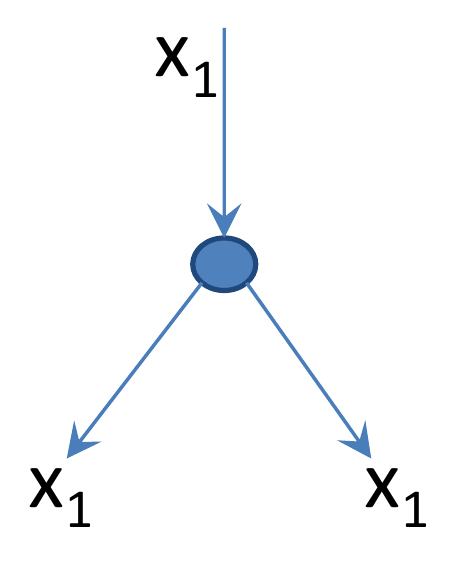}}
   \label{figure:broadcasting}}
   \caption{Coding operations defined on the nodes of $\graph'$.}
   \end{center}
\end{figure}

We find it desirable to work over a transformed graph ($\Vertices', \Edges'$) rather than the original graph ($\Vertices, \Edges$). This transformation can be done locally at each node, and results in a graph with some useful properties. In particular, we use the work of~\cite{LangbergSB:05} which demonstrates the equivalence between general network coding problems and those over ``low-degree" networks where each node has degree at most three. In particular, nodes in the reduced network either have one incident incoming edge and at most two incident outgoing edges (in which case they broadcast the incoming information on incident outgoing edges, and hence are called {\it broadcasting nodes}). Otherwise, they have two incident incoming edges and one incident outgoing edge (in which case they code the information on the incident incoming edges to generate information transmitted on the incident outgoing edge, and hence are called {\it coding nodes}). (See Figure~\ref{figure:broadcasting}.) This equivalence is useful for our probabilistic algorithms since it allows us to effectively enumerate networks. We change the equivalence relationship of~\cite{LangbergSB:05} slightly as described below so as to make it robust to nodes joining the original network. That is, in our equivalence relationship, nodes can join the original network while only locally perturbing the ``low-degree" network.

\begin{figure}[h]
   \begin{center}
   \scalebox{0.4}{\includegraphics{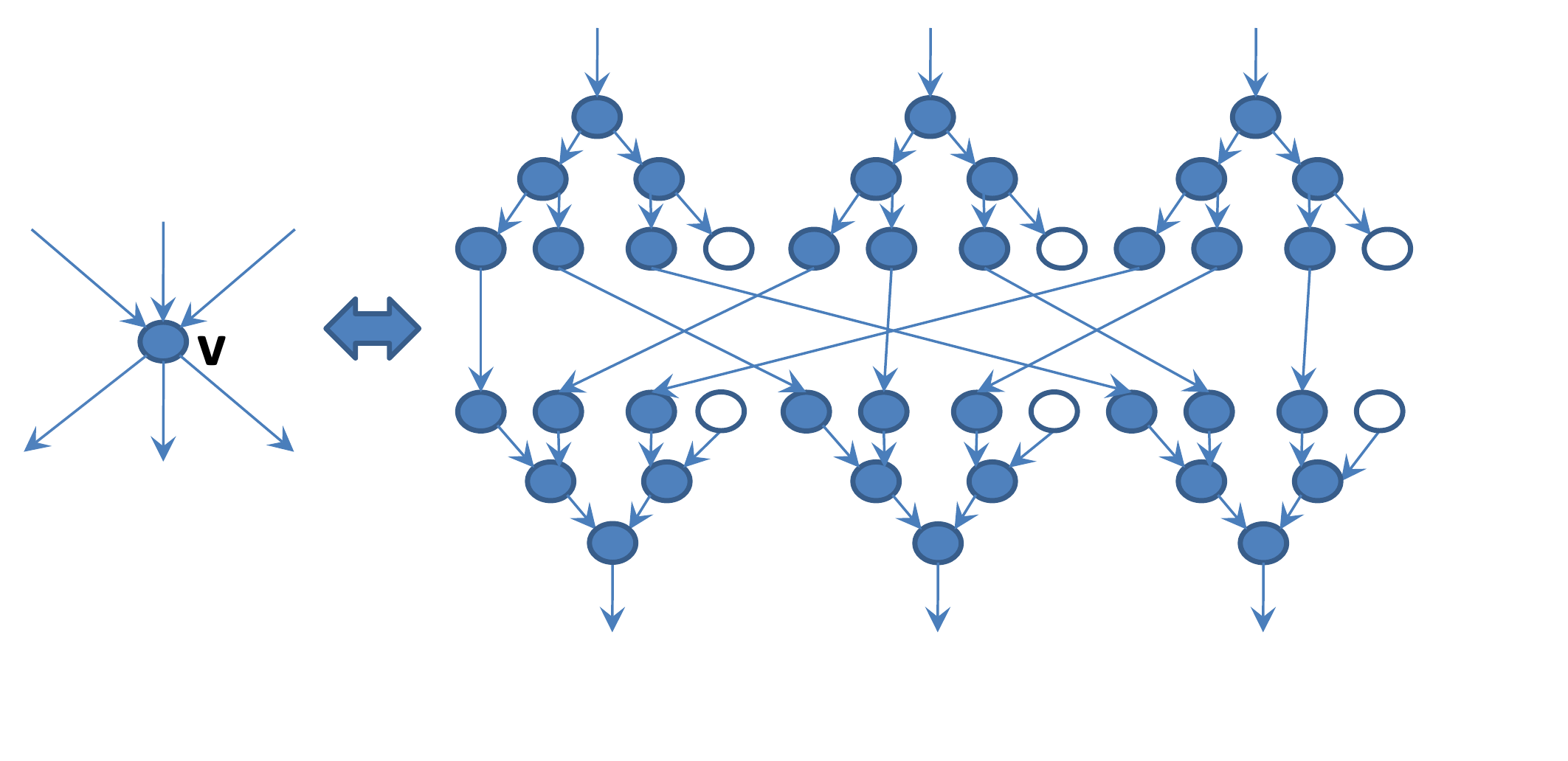}}
   \caption{A node $\nodev$ with three incoming and three outgoing links, and its corresponding virtual robust gadget.}
   \label{figure:gadget}
   \end{center}
\end{figure}

\begin{figure}[h]
   \begin{flushleft}
   \scalebox{0.4}{\includegraphics{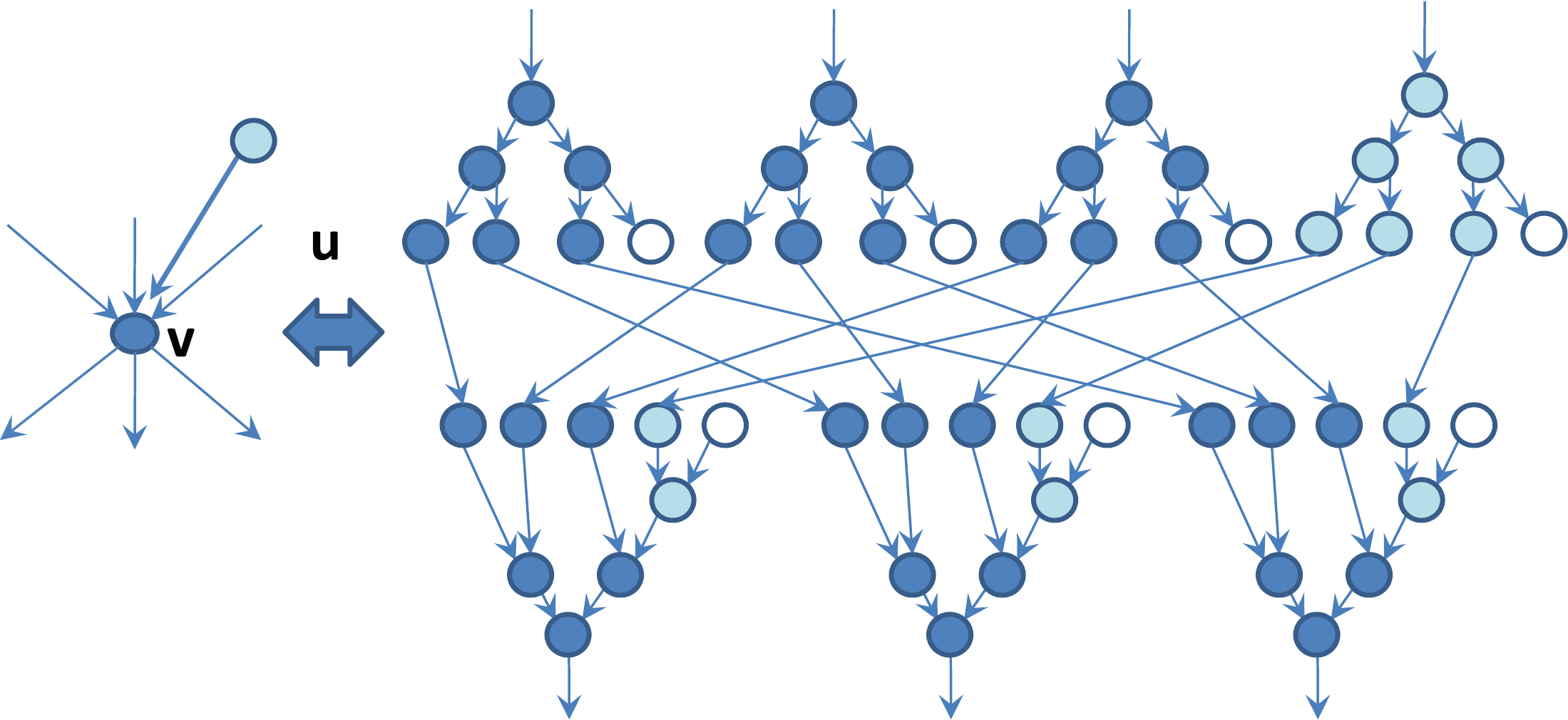}}
   \caption{Modification to the robust virtual gadget $G(v)$ of node $v$ when a new node $w$ connects to $v$.}
   \label{figure:new_node}
   \end{flushleft}
\end{figure}

The transformation is as follows. For every node $\nodev \in \Vertices$ we construct a {\it virtual robust gadget} $G(\nodev)$ (see Figure~\ref{figure:gadget} for an example\footnote{We thank Michael Langberg for providing the template for Figures~\ref{figure:gadget} and~\ref{figure:new_node}}).



Suppose $\nodev$ has $d_{in}(v)$ incoming links and $d_{out}(v)$ outgoing links. Corresponding each incoming link we construct a binary tree whose root is connected to that incoming link, and  which has $d_{out}(\nodev)+1$ leaves. Similarly, corresponding to each outgoing link we construct an inverted binary tree whose root is connected to that outgoing link, and which has $d_{in}(\nodev)+1$ leaves. The last leaf node of each binary tree is called a {\it virtual node}, and the other leaf nodes are called {\it connection nodes}. We then connect connection nodes so that there is exactly one path from each incoming link to each outgoing link (the connection order does not matter).

If a new link\footnote{A new node is treated as the corresponding set of new links created in the network. Similarly, a node's departure is treated as the set of corresponding links being removed.} is created in the network -- say, a link directed from $\nodew$ to $\nodev$ (see Figure~\ref{figure:new_node} for an example). In this case we first create a virtual gadget corresponding to the directed edge $(\nodew,\nodev)$. We then
split each virtual node on the inverted binary tree (corresponding to the outgoing links of $\nodev$) into two by appending a binary tree of depth one to it. We denote the second of the two new leaf nodes as a new virtual node, and the first as a new connection node. The connection nodes on $(\nodew,\nodev)$'s virtual gadget are then connected to the new connection nodes on each of the outgoing links' virtual gadgets so that there is exactly one path from $(\nodew,\nodev)$ to each link outgoing from $\nodev$. A corresponding (but inverted) procedure holds if the new link corresponded to a link {\it outgoing} from $\nodev$. The removal of a link simply corresponds to removal of the corresponding virtual gadgets on the incoming and outgoing sides, and all links connected to it.
The virtual nodes in each virtual gadget are what give our transformation robustness, since in case a new node joins or leaves the network, nodes other than the ones directly connecting with the changing node experience no structural changes in their existing virtual gadgets\footnote{The addition of virtual nodes and the corresponding robust connection procedure is the only substantive difference between our construction and that in~\cite{LangbergSB:05}}.

Henceforth, all algorithms in Section~\ref{sec:prob} shall convert the original graph to the virtual graph above as a pre-processing step, and all computations shall be over this virtual graph.
Also, as part of normal communication each node $\nodev$ in the virtual graph $(\Vertices',\Edges')$ estimates its {\it depth $\dep(\nodev)$}, {\it i.e.}, the length of the shortest path from the source to itself. This can be done by any of a variety of distributed shortest-path algorithms over acyclic graphs, such as the Bellman-Ford algorithm~\cite{CorLRS:01}.

\subsection{Weakly universal design}\label{subsec:weak}

{The essential idea behind our first scheme is as follows. 
Each node having estimated its depth, it then chooses a subset of $\F_2(z)$ whose size scales exponentially in this depth from which it pick its coding coefficients uniformly at random. We then show that the probability of error due to information being lost at any depth decays geometrically in the depth, and hence by the union bound the overall probability of error can be controlled so as not to exceed any desired $\epsilon$.}

\noindent {\bf \wupp (Weakly Universal Probabilistic) Code:} 
\begin{itemize}
\item Each coding node  $\nodev$ in the vertex-set $ \Vertices'$ of the virtual graph chooses two local coding coefficients corresponding to the two incoming links uniformly at random from the set of polynomials of degree at most $2\dep(\nodev)+1 + \log(\Rate|\sinks|/\epsilon)$.\footnote{This choice of the degree bound is simply to ease the analysis of Theorem~\ref{thm:wup}. All logarithms are binary. Also, for simplicity of presentation we assume that $\log(\Rate|\sinks|/\epsilon)$ is an integer -- if not, we may round up to the nearest integer with negligible error in our estimate of parameters.}

\end{itemize}
\begin{theorem}
For any $\epsilon > 0$, {\bf \wupp} has error probability at most $\epsilon$. 
\label{thm:wup}
\end{theorem}
\noindent {\bf Proof:}
Recall that $|\tran_\sink|$ represents the determinant of the transfer matrix from the source to sink $\sink$. As noted in~\cite{KoeM:02}, the network code is error-free if and only if the polynomial $\Pi_\sink|\tran_\sink|$ comprising of the product of the $|\tran_\sink|$ determinants over all sinks (with the network's local coding coefficients $\beta_{\nodeu,i,\nodev,j,\nodew}(z)$ as variables) is non-zero. To evaluate the probability that this is the case given the random assignment of local coding coefficients in {\bf \wupp}, we use Lemma~\ref{lem:SZ_Extend1}. Specifically, Each variable $x_i$ in Lemma~\ref{lem:SZ_Extend1} corresponds to a local coding coefficient. We group the coding coefficients $\beta_{\nodeu,i,\nodev,j,\nodew}(z)$ in terms of the depths $\dep(\nodev)$ of the nodes at which they are used. But there are at most $ 2^\dep \Rate$ coding nodes at any depth $\dep$ in the virtual graph, since after the transformation in Section~\ref{subsec:transform} the fast possible growth-rate for the new graph would be if it corresponded to $\Rate$ parallel binary trees -- one for each of the source's messages. Hence there are at most $2\Rate2^\dep$ local coding coefficients at that depth. 
Also, Corollary $1$ in~\cite{HoKMKE:03} shows that the degree of each local coding coefficient in $\Pi_\sink|\tran_\sink|$ is bounded from above by $|\sinks|$. 
By construction in \wupp each virtual node in the network chooses local coding coefficients uniformly at random from the set of polynomials of degree at most $2\dep(\nodev)+1 + \log(\Rate|\sinks|/\epsilon)$. This set is a subset of $\F_2(z)$ of size at most $2^{2\Rate\dep(\nodev)+1 + \log(|\sinks|/\epsilon)} = 2\Rate|\sinks|2^{2(\dep(\nodev)}/\epsilon$. 
Summing over all possible local coding coefficients at all possible (possibly infinite) depths and substituting the appropriate parameters into Lemma~\ref{lem:SZ_Extend1}, the error probability of the network code is bounded from above by
\begin{eqnarray*}
\Pr\left(\prod_{\sink \in \sinks}|\tran_\sink|)=0\right)
&\leq& \sum_{\dep = 1}^\infty 2\Rate2^\dep |\sinks|\frac{\epsilon}{2\Rate|\sinks|2^{2\dep}}\\
&=& \sum_{\dep = 1}^\infty  \frac{\epsilon}{2^\dep}\mbox{\hspace{0.2in}} =  \mbox{\hspace{0.2in}} \epsilon. \text{ \hspace{0.7in}} \Box
\end{eqnarray*}


The computational complexity of \wupp codes is polynomial in network parameters and $\log(1/\epsilon)$, and the achievable rates approach the network capacity $\Cp$ asymptotically in the block-length. Further, our codes are robust to links joining and leaving. Since the analysis of these properties is very similar to that of the codes in Section~\ref{subsec:sup}, we delay discussion to the end of that section.

\subsection{Strongly universal design}\label{subsec:sup}
We now present design of probabilistic robust linear network codes that are {\it strongly universal}, {\it i.e.}, independent of all network parameters. This obviates the requirement of knowledge of $\NumS$ of the codes in Section~\ref{subsec:weak}.
The idea underlying the construction in this section is as follows. For the purpose of analysis, for each sink we identify a set of edge-disjoint paths, and estimate the probability that the information on these edge-disjoint paths remains invertible as information flows through the network. In particular, for any sink $\sink$ and any depth $\dep$ in the network we identify the set of edges in these edge-disjoint paths that must contain linearly independent combinations of the source's information. {We call such sets of edges {\it flow-cuts}. It turns out that the number of flow-cuts} at any depth is in fact independent of the number of sinks, and further, a bound on this number at each depth can be computed locally. Thus sinks can be classified according flow-cuts. Hence, instead of trying to ensure that the linear transform to each sink is invertible as in Theorem~\ref{thm:wup}, nodes at each depth simply try to ensure that the linear transform to each flow-cut is invertible. To analyze the probability of non-invertibility at each flow-set an alternative to the end-to-end analysis of the probability of error used in~\cite{HoKMKE:03, JagCJ:03} is required. Here we use the proof technique of~\cite{JagCE:06}, which analyzes the probability that information gets lost from one set of edges in the network to a neighbouring set of edges. 

{\bf \supp (Strongly Universal Probabilistic) Code}
\begin{itemize}
\item {Each coding node $\nodev$ at depth $\dep(\nodev)$ in the vertex-set $ \Vertices'$ of the virtual graph chooses two local coding coefficients corresponding to the two incoming links uniformly at random from the set of polynomials of degree at most $(\Rate + 1)(\dep(\nodev) + 1 + \log\Rate)+\dep(\nodev) + \log(1/\epsilon)-1$}.
\end{itemize}   
\begin{figure}[h]
   \begin{center}
   \subfigure[Butterfly network.]
   {\scalebox{0.4}{\includegraphics{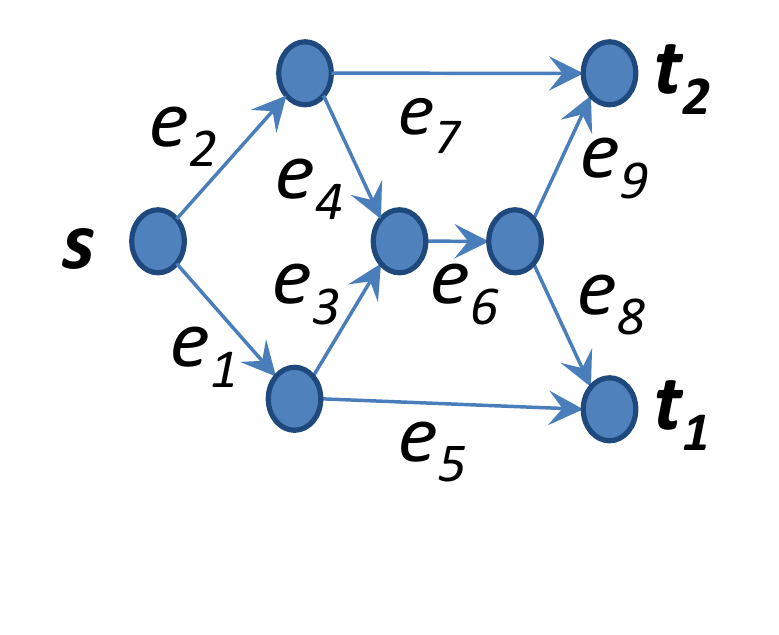}}
   \label{figure:butterfly}}
   \subfigure[One possible choice of a flow-set $\flowset({\sink}_1)$ for the butterfly network.]
   {\scalebox{0.4}{\includegraphics{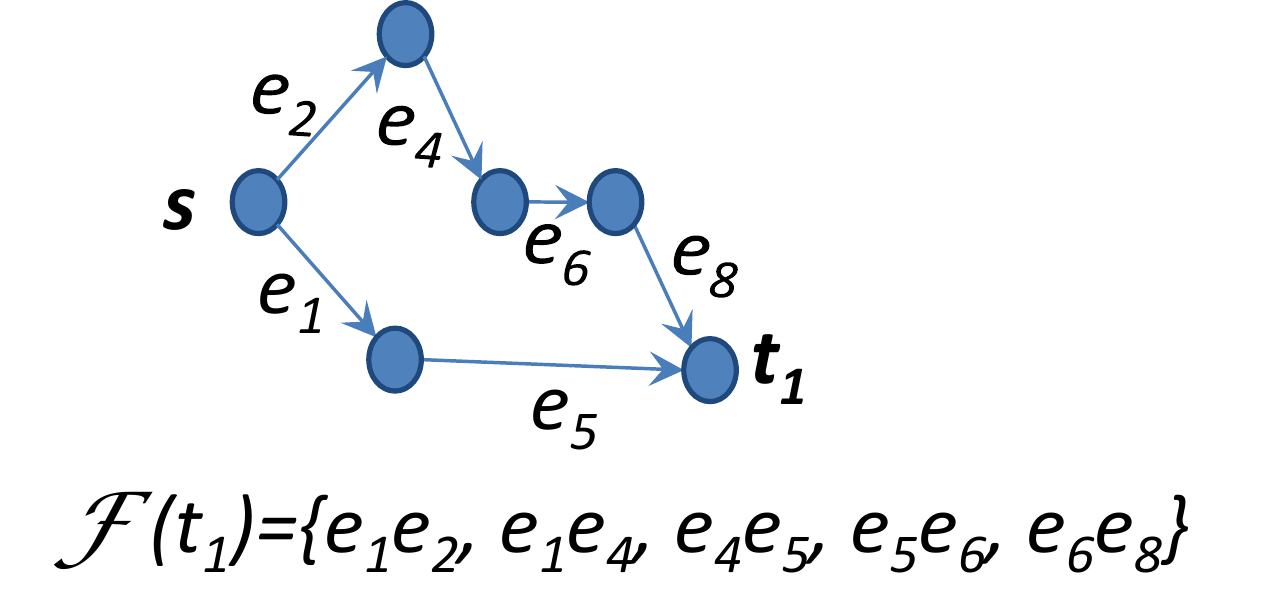}}
   \label{figure:flowcut}}
   \caption{Illustration of the definition of a flow-cut and a flow-set based on the butterfly network topology. The flow-set $\flowset(\sink_1)$ comprises of the successive flow-sets $(\edge_1,\edge_2)$, $(\edge_1,\edge_4)$, $(\edge_5,\edge_4)$, $(\edge_5,\edge_6)$, $(\edge_8,\edge_6)$.
}
   \end{center}
\end{figure}
Recall that by assumption the capacity of the network is at least $\Rate$. Hence there is a set $\cP$ of at least $\Rate$ edge-disjoint paths 
that go from the source $\source$ to each $\sink$. 

Corresponding to each such set $\cP$ of edge-disjoint paths, 
we define {\it flow-cuts}. A {\it flow-cut} $\flowcut(\sink)$ is defined as a set of $\Rate$ edges that have the property that each edge in the flow-cut is from a distinct edge-disjoint path in $\cP(\sink)$. These flow-cuts are useful since we intend to analyze the linear (in)dependence of information flowing through the edges in each flow-cut -- if the information on each edge in a flow-cut is linearly independent, then the source's information can be retrieved from that flow-cut. Hence, we only need to inductively prove that no information is lost from one flow-cut to the ``next" flow-cut, appropriately defined as below\footnote{Similar intuition was used in the proofs of~\cite{JagSanCEEJT} and ~\cite{JagCE:06}, where they were called ``frontier edge-sets".{Note that a flow-cut need not be a cut or a subset of it -- for instance, it may include two edges on two edge-disjoint paths, such that one is incoming to a node, and the other is outgoing from it.}}.

We define the {\it depth $\dep(\flowcut^\sink)$ of a flow-cut} $\flowcut^\sink$ as the maximum depth of the head of any edge in it, {\it i.e.}, $\dep(\flowcut(\sink))=\displaystyle \max_{head(\edge) \in \flowcut(\sink)} \dep(e_i)$. Further, we denote a flow-cut of depth $\dep$ by $\flowcut(\sink, \dep)$. 

We then define a {\it flow-set} $\flowset(\sink)$ as an ordered set of flow-cuts with some properties. In particular, each flow-cut in a flow-set differs from the successive flow-cut in exactly one edge. Specifically, if one flow-cut in $\flowset(\sink)$ differs from the next flow-cut in $\flowset(\sink)$ in that some edge $\edge$ is replaced by another $\edge'$, then it must be the case that $\edge$ is the edge preceding $\edge'$ in some path in the set of edge-disjoint paths $\cP(\sink)$. Intuitively, each flow-set $\flowset(\sink)$ captures successive snapshots of how information flow from the source to the sink $\sink$.

Examples of flow-cuts and flow-sets are provided in Figure~~\ref{figure:flowcut}, based on the butterfly network in Figure~\ref{figure:butterfly}.


Let $\flowcut(\sink,\dep)$ be some flow-cut of depth $\dep$ to sink $\sink$, and $\flowcut'(\sink)$ be the flow-cut immediately preceding\footnote{Note that the depth of $\flowcut'(\sink)$ might be either $\dep$ or $\dep-1$, since two successive flow-cuts differ in exactly one edge, which may or may not be the deepest edge in a flow-cut (if not, then both flow-cuts have the same depth; if so, the depth of the flow-cut can change by at most one).} $\flowcut(\sink,\dep)$ in flow-set $\flowset(\sink)$.
Let $\tran(\flowcut(\sink,\dep))$ be the linear transform that the network imposes from the source $\source$ to the edges in the flow-cut $\flowcut(\sink,\dep)$, and let $\rho(\dep)$ be the rank of this transform. Correspondingly, let $\tran(\flowcut'(\sink))$ be the linear transform from $\source$ to $\flowcut'(\sink)$, and let $\rho'$ be the rank of this transform. 
Then the following lemma gives an upper bound on the probability that choosing local coding coefficients according to the dictates of \wupp results in a loss of information in going from $\flowcut'(\sink)$ to $\flowcut(\sink,\dep)$.
\begin{theorem}
\label{thm:sup}
For every $\epsilon>0$, \supp has error probability at most $\epsilon$.
\end{theorem}
\noindent  {\bf Proof:}
Note that of the two types of nodes in the virtual graph, the broadcasting nodes induce no additional error -- if a flow-cut contains $\Rate$ linearly independent packets, and one of the edges in the flow-cut is replaced with another edge at a broadcasting node, the information in the succeeding flow-cut remains unchanged. Thus from now on we focus only on coding nodes.

By construction the structure of the virtual graph $(\Vertices',\Edges')$ is such that each node can have at most two outgoing edges, and further the source node is replicated $\Rate$ times. Hence the maximum possible number of edges in the virtual graph up to a depth $\dep$ occurs when it comprises of $\Rate$ parallel binary trees. But each binary tree has at most $2(2^\dep)$ edges, hence the total number of edges in the virtual graph up to depth $\dep$ is at most $2\Rate(2^\dep)$. Also, the total number of flow-sets of depth $\dep$ is at most ${2\Rate(2^\dep)}\choose{\Rate}$, which is bounded from above\footnote{{Since ${{a}\choose{b}} \leq a^b$. Also, all exponents $exp$ are base $2$.}} by $exp(\Rate(\dep + 1 + \log\Rate))$.

We use these bounds to bound from above the number of distinct type of coding choices a coding node at a certain depth faces. All our analysis now focuses on the specific following coding node $\nodev$. Let its incoming edges be $\edge'$ and $\edge''$, and the outgoing edge be $\edge$. Let edge $\edge'$ belong to a flow-cut $\flowcut'(\sink)$ in flow-set $\flowset(\sink)$ going towards sink $\sink$, and edge $\edge''$ be an arbitrary other edge. Then the outgoing edge $\edge$ replaces $\edge'$ in the flow-cut $\flowcut'(\sink)$ to produce flow-cut $\flowcut(\sink)$. Suppose $\flowcut(\sink)$ is of depth $\dep$.
Then by the bounds in the preceding paragraph, the number of ways an arbitrary flow-cut of depth $\dep$ can result from the merger of a preceding flow-cut and an arbitrary edge of depth at most $\dep$ is at most $2\Rate(exp(\dep)) \times exp(\Rate(\dep + 1 + \log\Rate))$, which equals 
\begin{equation}
exp((\Rate + 1)(\dep + 1 + \log\Rate)).
\label{eq:bnd_mix}
\end{equation}

Next, we estimate the probability that a coding node ``loses information". That is, we bound from above the probability that the number of linearly independent packets on the edges of a flow-cut is less than $\Rate$ even though the immediately preceding flow-cut has $\Rate$ linearly independent packets.

Say $\tran(\flowcut(\sink,\dep))$ represents the $\Rate \times \Rate$ matrix whose $i$th row represents the linear combinations of the source's $\Rate$ messages on the $i$th link in the flow-cut $\flowcut(\sink,\dep)$ of depth $\dep$. Correspondingly, let $\tran(\flowcut'(\sink))$ represent the matrix representing the linear transform from the source to the flow-cut $\flowcut'(\sink)$ immediately preceding $\flowcut(\sink,\dep)$ in the flow-set $\flowset(\sink)$, and suppose it is of full rank $\Rate$. Then the message $\eY_\edge(z)$ on edge $\edge$ in flow-cut $\flowcut(\sink,\dep)$ may be written as $\beta_{\edge',\edge}(z)\eY_{\edge'}(z) + \beta_{\edge'',\edge}(z)\eY_{\edge''}(z)$. (Recall that $\edge'$ and $\edge''$ are the edges incoming to $\nodev$, $\eY_{\edge'}(z)$ and $\eY_{\edge''}(z)$ are the corresponding messages carried by them, and $\beta_{\edge',\edge}(z)$ and $\beta_{\edge'',\edge}(z)$ represent the local coding coefficients at node $\nodev$.) But by assumption $\tran(\flowcut'(\sink))$ is of full-rank, and hence the message $\eY_{\edge''}(z)$ may be written as a linear combination of the messages on the edges in flow-set $\flowset'(\sink)$. Thus the message on edge $\edge$ may be written as
$$\beta_{\edge',\edge}(z)\eY_{\edge'}(z) + \sum_{\edge(i)\in \flowcut'(\sink)} \gamma_{\edge(i),\edge}(z)\eY_{\edge''}(z),$$
\noindent for some $\gamma_{\edge(i),\edge}(z) \in \F_2(z)$. This in turn equals
$$(\beta_{\edge',\edge}(z)+\gamma_{\edge'',\edge}(z))\eY_{\edge'}(z) + \sum_{\edge(i)\in \flowcut'(\sink): \edge(i)\neq \edge'} \gamma_{\edge(i),\edge}(z)\eY_{\edge''}(z).
$$
But the information on the links in $\flowset(\sink,\dep)$ other than $\edge$ is unchanged, and hence the only manner in which the messages on the edges in $\flowset(\sink,\dep)$ are linearly dependent is if the coefficient $(\beta_{\edge',\edge}(z)+\gamma_{\edge'',\edge}(z))$ equals zero.
{But by the choice specified in \supp the coding coefficients are chosen from the set of polynomials of degree at most $\left ( -\left [ (\Rate + 1)(\dep + 1 + \log\Rate)+\dep + \log(1/\epsilon - 1)\right ] \right )$ -- this set is of size $exp\left ( -\left [ (\Rate + 1)(\dep + 1 + \log\Rate)+\dep + \log(1/\epsilon)\right ] \right )$.
Lemma~\ref{lem:SZ_Extend1} then implies that the probability that the degree $1$ polynomial $(\beta_{\edge',\edge}(z)+\gamma_{\edge'',\edge}(z))$ equals zero, is at most $exp\left ( -\left [ (\Rate + 1)(\dep + 1 + \log\Rate)+\dep + \log(1/\epsilon)\right ] \right )$. Analogously to Theorem~\ref{thm:wup}, taking the union bound over all possible coding operations at depth $\dep$ (for which (\ref{eq:bnd_mix}) is an upper bound), and
summing over all (possibly infinite) depths $\dep$ gives us that the overall probability of error is at most
\begin{eqnarray*}
\sum_{\dep = 1}^\infty 2^{(\Rate + 1)(\dep + 1 + \log\Rate)}\frac{\epsilon}{2^{\dep +\left ( \left [ (\Rate + 1)(\dep + 1 + \log\Rate)\right ] \right )}}\\
= \sum_{\dep = 1}^\infty  \frac{\epsilon}{2^\dep} \mbox{\hspace{0.2in}} =  \mbox{\hspace{0.2in}}\epsilon. \text{ \hspace{1.4in}} \Box
\end{eqnarray*}
}

\subsection{Robustness}
{ Due to the robust graph transformation described in Section~\ref{subsec:transform}, neither the addition nor the deletion of edges or nodes in the network causes problems with our proof. If new nodes are added to the network, the actual depth of some nodes in the network, say $\nodev$ in particular, may decrease. However, we require each node $\nodev$ to use in perpetuity the value of the depth $\dep(\nodev)$ it estimates in the first round of communication. This ensures that the bound on the number of coding nodes at a particular depth $\dep$ is not violated. Conversely, if nodes leave the network, the actual depth of $\nodev$ may increase. Nonetheless, the bound on the number of coding nodes at a particular depth $\dep$ is still not violated, since the total number of nodes in a network has only decreased.}

\subsection{Complexity analysis}
\label{prob_complexity}
{The complexity of both \wupp and \supp scale with the corresponding degree of the polynomials chosen by nodes as coding coefficients.
 
For \wupp, by construction, the degree of the polynomials used as coding coefficients (as noted in~\cite{JagEHM:04} is a good proxy for the implementation complexity of codes) scales as the maximum depth of the virtual graph plus ${\mathcal O}(\log(\Rate\NumS/\epsilon))$. 
But in the virtual graph each original node is replaced with a gadget with depth that is approximately the logarithm of the degree of the node, which in the worst case is at most\footnote{Recall that $\maxcap$ denotes the maximum value for any edge's capacity in the network.} $\maxcap\NumV$. Hence the maximum depth of the virtual graph is at most ${\mathcal O}(\log(\maxcap\NumV))$ times the number of vertices in the original graph. Pulling it all together, the complexity of implementation scales as ${\mathcal O}(\NumV\log(\maxcap\NumV) + \log(\Rate\NumS/\epsilon))$.

The corresponding redundancy the network introduces in the codes, arising from the delays introduced by each coding node, then scales at worst as the maximum depth of the virtual graph times the maximum degree of the polynomials used by any node, since each coding node in a path can introduce at most the maximal delay and delays along a path add up. Using the bounds above, this scales as ${\mathcal O}(\NumV^2\log^2(\maxcap\NumV) + \NumV\log(\maxcap\NumV)\log(\Rate\NumS/\epsilon))$.
%

A similar analysis shows that the complexity of implementation of \supp scales as ${\mathcal O}(\Rate\NumV^2\log^2(\maxcap\NumV) + \NumV\log^2(\maxcap\NumV) \log(1/\epsilon)\})$, and that the redundancy introduced by such codes scales as ${\mathcal O}((\Rate\NumV^3\log^3(\maxcap\NumV) + \NumV^2\log^3(\maxcap\NumV) \log(1/\epsilon)\}))$.
}


We now demonstrate that the implementation complexity of \wupp and \supp is in fact order-optimal by demonstrating a class of networks for why any universal design requires computational complexity which is similar in order of magnitude to that of \wupp and \supp. Our construction is inspired by that of~\cite{LL04}.
Consider any universal $\epsilon$-error linear network code, {\it i.e.}, any network code that requires that each sink be able to reconstruct the source's information with probability of error at most $\epsilon$, even if the network topology is not known in advance.
\begin{theorem}
There exists a class of networks for which the implementation complexity for any $\epsilon$-error universal network code is $\Omega({\NumE} - \log(1/\epsilon))$.
\label{thm:lower_bound}
\end{theorem}
\noindent {\bf Proof: }
\begin{figure}[h]
   \begin{center}
   {\scalebox{0.50}{\includegraphics{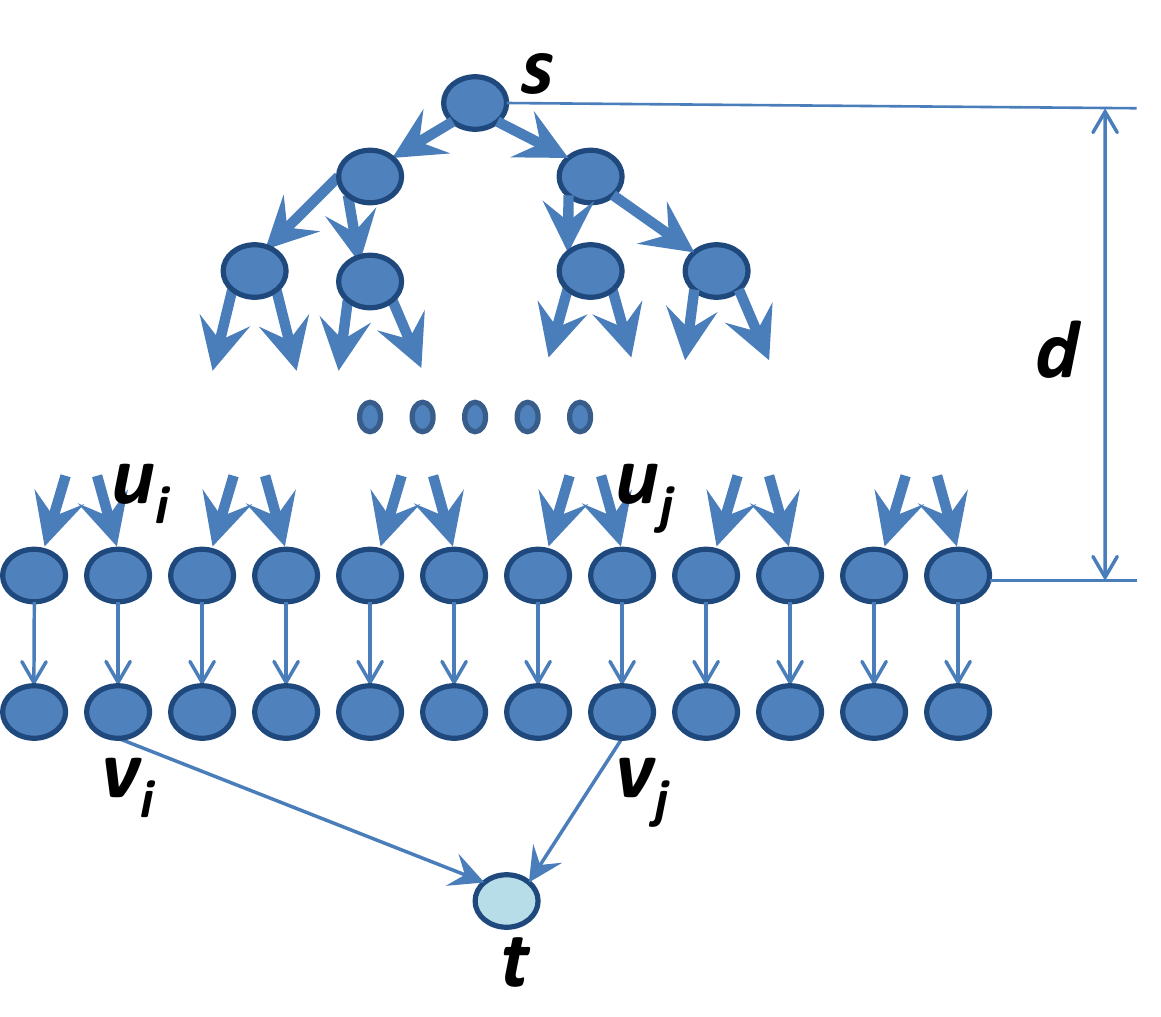}}
   \label{figure:example_topology_compl}}
   \caption{Example demonstrating that the implementation complexities of \wupp and \supp are order-optimal.}
   \end{center}
\end{figure}
We construct a single-source multicast network that requires that for any universal code, coding operations must be chosen from a set that must be exponentially larger than would actually be needed if the topology were known in advance. Consider the ``binary-tree-like" network demonstrated in Figure~\ref{figure:example_topology_compl}. The upper part of this graph comprises of a binary tree of depth $\dep$, with the source $\source$ located at the root of the tree, and hence $2^\dep$ leaf nodes. Each link of this binary tree has capacity $2$. Next, each leaf node of this binary tree has a link of unit capacity leaving it to a corresponding {\it forwarding node}. Finally, we consider two possibilities. Either there are ${2^\dep}\choose{2}$ sinks, such that each sink is connected to a distinct subset of size two of the set of ${2^\dep}$ forwarding nodes, via unit capacity links to each of the two nodes in the subset (this is in the spirit of the {\it combination networks} examined in~\cite{LL04}).
Or, there is only one sink node in the network, which is connected to two of these forwarding nodes (say $\nodev_i$ and $\nodev_j$) via links of unit capacity each.

As to coding strategies, each node in the binary tree part of the network can forward two linearly independent messages $X_1(z)$ and $X_2(z)$ on each of its outgoing links. Hence at depth $\dep$ each of the $2^\dep$ leaves of the binary tree have both $X_1(z)$ and $X_2(z)$. 
Since neither of the two leaf node corresponding to $\nodev_i$ or $\nodev_j$ (say we call them $\nodeu_i$ and $\nodeu_j$ respectively) is aware of which of the two configurations the network is in, to be universal they must use coding operations that work for both configurations.
 
First we consider the case of $\epsilon = 0$, {\it i.e.}, every message must be decoded correctly.
Suppose the leaf nodes $\nodeu_i$ and $\nodeu_j$ choose to transmit the linear combinations $A_1(z)X_1(z)+B_1(z)X_2(z)$ and $A_2(z)X_1(z)+B_2(z)X_2(z)$, or equivalently $X_1(z)+\alpha_1(z)X_2(z)$ and $X_1(z)+\alpha_2(z)X_2(z)$ (by setting $\alpha_i(z) = B_i(z)/A_i(z)$ for $i = 1,2$). But the messages on each of the forwarding links must be linearly independent (to take into account the eventuality that the network is in the first configuration with ${2^\dep}\choose{2}$ sinks). Hence there must be at least $2^\dep - 1$ choices for the $\alpha_i(z)$s (one for each of the leaf nodes, minus one for the case when $A_i(z) = 0$, which can be handled separately.) 
But if the set of possible coding operations is $\Omega(2^\dep)$, then its implementation complexity must be at least $\Omega(\dep)$. But this is $\Omega(\NumE)$. In contrast, if the network happened to be in the configuration with only one sink and this was known in advance, then each of $\nodeu_i$ and $\nodeu_j$ could simply forward one bit, for an implementation complexity of $1$.

The case with $\epsilon$-errors can be similarly analyzed, by allowing sinks to make errors a fraction $\epsilon$ of the time. A direct counting argument gives the required result.
\hfill $\Box$

\section{Deterministic designs}\label{sec:starwars}

In this section we describe two deterministic designs of universal distributed robust network codes that are zero-error\footnote{Preliminary versions of the proofs in this section were in the thesis~\cite{Jaggi:06}}.

Our first scheme is only for codes of rate $2$. It is related to a construction of~\cite{FraS:04}, but generalizes it so that the choice of coding operations is independent of the size of the network. We call our scheme the {\bf Rate $2$ Deterministic Design} \rtdt for short.
Our purpose in presenting this first scheme is primarily expository, since the proof is significantly easier than that of the second scheme -- it helps set the stage for the second scheme.

Our second scheme is for general rates and is independent of all network parameters, including the number of sinks. We call this scheme the {\bf Capacity $3$ or more, Probability of error $0$} scheme, or \ctpo for short.
However, \ctpo is more of an existence result than a practical code since the computational complexity of its implementation is exponential in network parameters.

We first describe some useful preprocessing steps relevant for both of our schemes.

\subsection{Robust distributed unique ID assignment}
\label{Robust_ID}
While the codes in Section~\ref{sec:prob} only required nodes to estimated their depth, the zero-error codes in presented in this section require nodes to obtain a {\it unique ID}, {\it i.e.}, an ID that is distinct for each node in the network. Such an ID allows nodes to loosely coordinate coding choices even if they are unable to communicate directly with each other, and thereby ensure that the overall code is ``good". Such IDs might be pre-assigned to nodes (for example via factory stamps, or GPS coordinates, or IP addresses), or be assigned on the fly, as described below.

The task of distributing unique IDs to nodes over a directed graph was considered in~\cite{BruLS:07}. The essential idea of their algorithm is to pretend that the graph is a tree directed from the root to the leaves (if not, extra edges are removed for the ID assignment protocol), and to assign IDs so that the binary expansion of each node's ID is a prefix to the binary expansion of all nodes downstream from it. 
This ID distribution can be carried out with communication cost that is asymptotically negligible in the packet length, in conjunction with the normal flow of information through the network, for instance in the header. 
Here, as in Section~\ref{subsec:transform}, we need to change the unique ID distribution protocol slightly to make it robust to network changes, so that new nodes are still ensured that IDs assigned to them do not clash with previously assigned IDs. In the same spirit as the robust virtual gadgets in Section~\ref{subsec:transform}, at each node $\nodev$ we reserve a {\it virtual ID} for the event that a new node might in the future connect to $\nodev$; if so, this virtual ID is again split into another virtual ID, and an ID that is assigned to the new node.
{As noted in~\cite{BruLS:07} the worst-case growth rate of the largest node ID with the network size is exponential in $\NumV$, for reasons similar to those outlined in Theorem~\ref{thm:lower_bound} -- nodes might be unable to distinguish between a full binary tree, and a very sparse graph.}

\subsection{Cantor labeling}
\label{C_Graph}
The well-known Cantor diagonal argument~\cite{Cantor:1874} makes an unexpected cameo in this work. One version states that the cardinality of the set of integers is the same as that of the set of finite dimensional vectors with integer components, and further gives an effective bijection between the two sets. Further, this bijection guarantees that any vector in ${\mathbb Z}^k$ with maximum component $l$ is mapped to an integer of size $\theta(l^k)$. This mapping is useful since, given a unique ID for each node $\nodev$, we then need to produce unique coding coefficients for each pair of edges such that one is incoming to $\nodev$ and the other is outgoing for $\nodev$. Prior to code design, the number of such coefficients that each node might need to choose is unknown. However, each $\beta_{\nodeu,i,\nodev,j,\nodew}(z)$ coefficient can be labeled by at most the five indices $(\nodeu,i,\nodev,j,\nodew)$, each of which is an integer. Hence given a node's unique ID, one can produce unique integral labels for each vector $(\nodeu,i,\nodev,j,\nodew)$ that are not too much larger (at most the fifth power) than any of the five parameters in $(\nodeu,i,\nodev,j,\nodew)$. This mapping, denoted $\cantor(.)$,  can then be used to select distinct local coding coefficients as needed in Sections~\ref{subsec:r2d2} and~\ref{subsec:c3p0}

\begin{figure}[h]
   \begin{center}
   {\scalebox{0.60}{\includegraphics{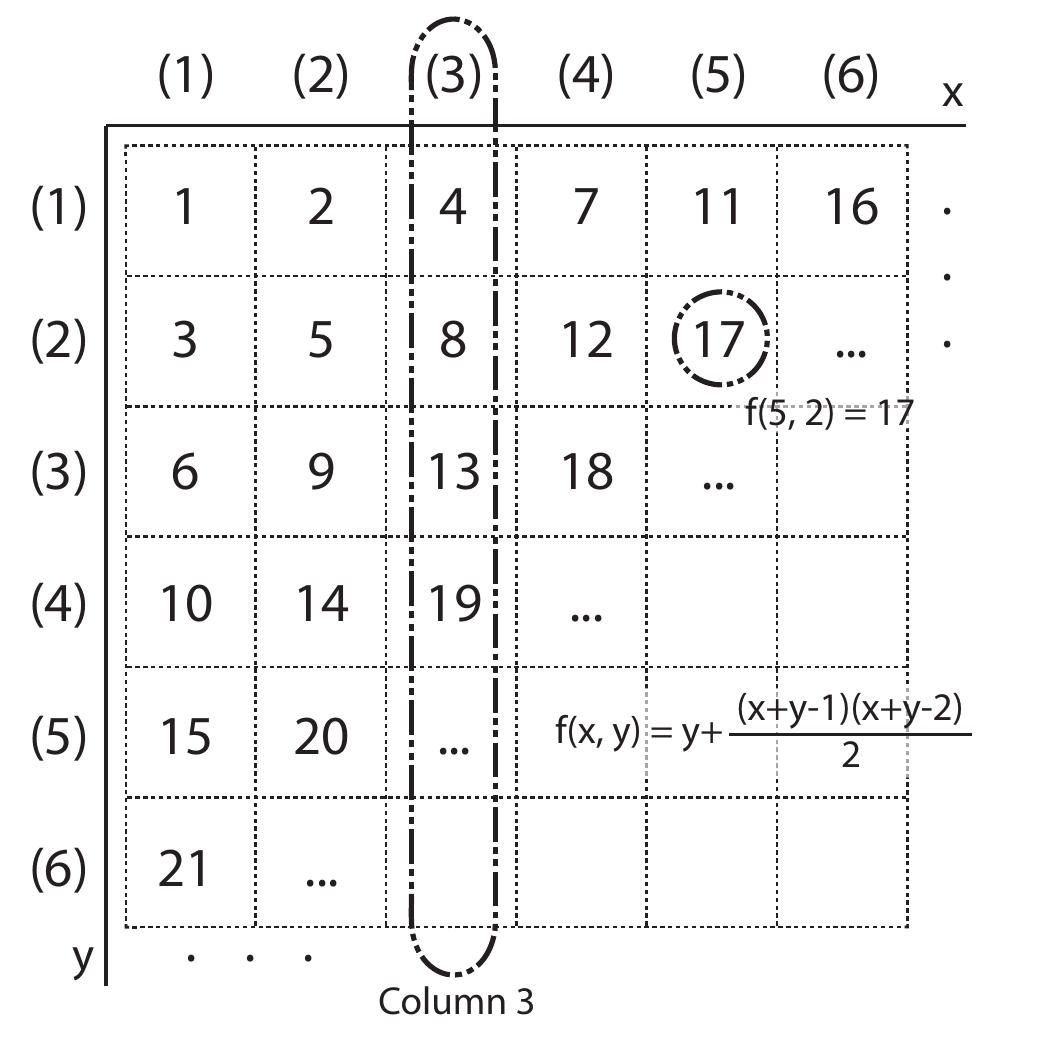}}
   \label{figure:zero_error}}
   \caption{Example showing the mapping between ${\mathbb Z}$ and ${\mathbb Z}^2$.}
   \end{center}
\end{figure}




\subsection{Rate $2$ zero-error codes}\label{subsec:r2d2}
For the case when the transmission rate equals $2$, note that there are essentially just two non-trivial scenarios for each node -- either a node receives one linearly independent message on incoming links, or it receives two. In the former case, it can only broadcast incoming information on outgoing links. In the latter case, it can reconstruct the source's information, and thereby can fully control the linear combinations on outgoing links. Our construction  for \rtdt rests on analysis of these cases.

\noindent {\bf \rtdt (Rate $2$ Deterministic Design)}
\begin{itemize}
\item The source $\source$ has two linearly independent messages $X_1(z)$ and $X_2(z))$.
\item
Depending on its connectivity to the source, on incoming edges each node $\nodev \in \Vertices$ receives either one or two linearly independent combinations of the source messages $(X_1(z),X_2(z))$.
\item 
If a node $\nodev$ receives only one linearly independent message on incoming links, it broadcasts it down all outgoing edges.
\item
If a node $\nodev$ receives two linearly independent combinations of
$(X_1(z),X_2(z))$, this enables it to reconstruct both $X_1(z)$ and $X_2(z)$. For each $j$th directed edge connecting each pair of nodes $\nodev$, $\nodew$ (connected possibly by multiple parallel edges), we use the Cantor labeling algorithm\footnote{In Section~\ref{C_Graph} we assume that $\nodeu$ and $i$ are also variable, but in this section they are fixed.} in Section~\ref{C_Graph} to assign a distinct local coding coefficient. In particular, let $\cantor(\nodev,j,\nodew)$ denote the $3$-dimensional Cantor mapping. Then the node $\nodev$ then transmits $X_1(z) + \beta_{\cantor(\nodev,j,\nodew)}(z)X_2(z)$ down the $j$th edge connecting $\nodev$ to $\nodew$ (here $\beta_{\cantor(\nodev,j,\nodew)}(z)$ is chosen to be distinct for each $\cantor(\nodev,j,\nodew)$).
\end{itemize}
\begin{theorem}
For any network $\graph$ with min-cut capacity at least $2$, \rtdt succeeds with zero error. 
\label{thm:r2d2}
\end{theorem}
\noindent {\bf Proof:}
For any $\nodev \in \Vertices$ such that the mincut between the source and $\nodev$ is at least $2$, there are at least two edge-disjoint paths from the source to $\nodev$. By the statement of our \rtdt algorithm, for any such nodes $\nodev$ and $\nodev'$, the linear combinations of $X_1(z)$ and $X_2(z)$ on all their outgoing links must be distinct, and linearly independent (since the vectors $(1, \beta_{\cantor(\nodev,j,\nodew)}(z))$ and $(1, \beta_{\cantor(\nodev',j',\nodew')}(z))$ are linearly independent if and only if $\beta_{\cantor(\nodev,j,\nodew)}(z)$ and $\beta_{\cantor(\nodev',j',\nodew')}(z)$ are distinct). 
\hfill $\Box$

\subsection{General zero-error codes} \label{subsec:c3p0}
The challenge in extending the results of Section~\ref{subsec:r2d2} to rates greater than $2$ lies in the fact that there might be nodes receiving two or more linearly independent pieces of information, and yet are unable to decode the source messages. In this case, they do not have full control over the messages they are able to send out, and hence the argument of Theorem~\ref{thm:r2d2} fails. In this section, we get around this challenge by examining a different invariant of linear convolutional network codes. In particular, we choose coding coefficients in a distributed manner so that the delay of the source messages on {\it every} path in the network is distinct. This means that the source messages never cancel out at the sinks, and hence can be reconstructed.
\begin{figure}[h]
   \begin{center}
   {\scalebox{0.35}{\includegraphics{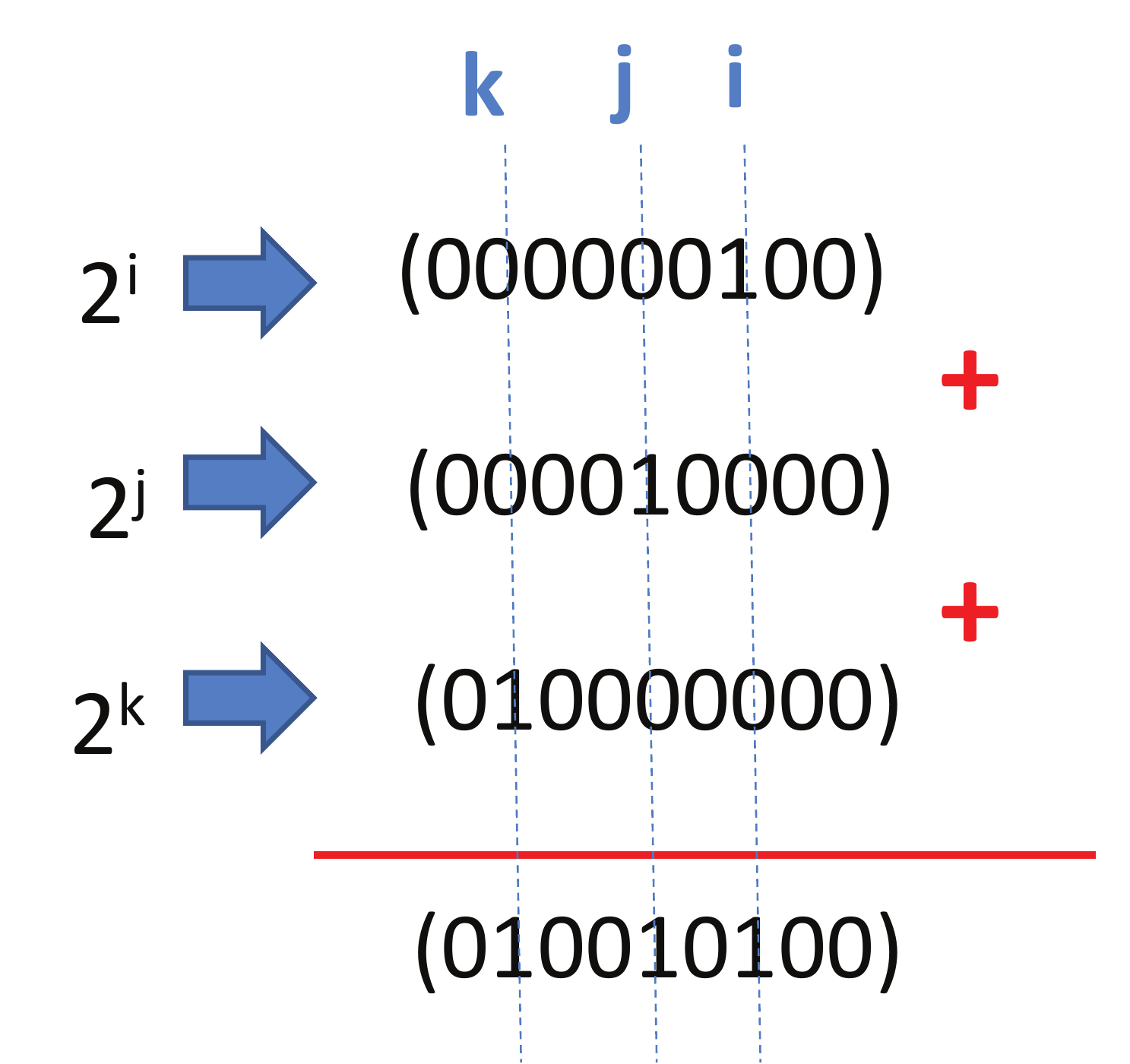}}
   \label{figure:zero_error}}
   \caption{Graphic representation of the proof of correctness of  \ctpo codes.}
   \end{center}
\end{figure}

\noindent {\bf \ctpo (Capacity $3$ or more, Probability of error $0$) codes}
\begin{itemize}
\item For each $5$-tuple $(\nodeu,i,\nodev,j,\nodew)$, let $\cantor(\nodeu,i,\nodev,j,\nodew)$ be the $5$-dimensional Cantor mapping defined in Section~\ref{C_Graph}. 
We define the local coding coefficient $\beta_{\nodeu,i,\nodev,j,\nodew}(z)$ as
\begin{eqnarray*}
\label{coef_assignment}
z^{2^{\cantor(\nodeu,i,\nodev,j,\nodew)}},
\end{eqnarray*}
\noindent {\it i.e.}, the monomial in $z$ with degree $exp({\cantor(\nodeu,i,\nodev,j,\nodew)})$ (here the exponent is base $2$).
\end{itemize}
\begin{theorem}
For any network $\graph$, \ctpo succeeds with zero error. 
\end{theorem}
\noindent {\bf Proof:}
Theorem $4$ in~\cite{HoKMKE:03} demonstrates that $|\tran_\sink|$, the determinant of the transfer matrix $\tran_\sink$ from the source to any sink $\sink$, can be written as $\Sigma c_\cP \Pi_\cP \beta_{\nodeu,i,\nodev,j,\nodew}(z)$, where the product is over all the local coding coefficients on a particular path $\cP$ from $\source$ to $\sink$, $c_\cP$ is a non-zero constant corresponding to $\cP$, and the outer summation is over all paths from $\source$ to $\sink$. Our choice of local coding coefficients along any path in \ctpo implies that $|\tran_\sink|$ equals
\begin{equation}
\sum c_\cP \prod_\cP z^{exp({\cantor(\nodeu,i,\nodev,j,\nodew)})} = \sum c_\cP z^{(\sum_\cP exp({\cantor(\nodeu,i,\nodev,j,\nodew))})}.
\label{eq:ctpo}
\end{equation}
But by choice, each of the terms $\cantor(\nodeu,i,\nodev,j,\nodew)$ is distinct, and hence the binary expansion of $exp(\cantor(\nodeu,i,\nodev,j,\nodew))$ has a single $1$ in a distinct location. But if two paths in the summation (\ref{eq:ctpo}) differ, then they must differ in at least one of the local coding coefficients, and therefore the exponent of the power of $z$ along the two paths must differ -- hence each path corresponds to a distinct power of $z$. This implies that as long as there is at least one path from $\source$ to each $\sink \in \sinks$, each of the corresponding transforms $\tran_\sink$ must be invertible. 
\hfill $\Box$

\subsection{Complexity Analysis}
{The complexity of both \rtdt and \ctpo scale with the corresponding Cantor labeling and node assignments. 

For \rtdt the size of the set any node chooses its coding coefficients from scales as the third power of the largest node ID or the largest link-capacity in the network. But as noted in Section~\ref{C_Graph}, the largest node ID can scale exponentially in $\NumV$. Hence the degree of the polynomials used as coding coefficients scales logarithmically in the size of the sets from which local coding coefficients are chosen, which in turn scales as ${\mathcal O}(\max\{\NumV,\log^3(\maxcap)\})$. The corresponding redundancy the network introduces in the codes, arising from the delays introduced by each coding node, then scales as ${\mathcal O}(\NumV \max\{\NumV,\log^3(\maxcap)\})$, since each coding node in a path can introduce at most the maximal delay and delays along a path add up. 

A similar analysis shows that the complexity of implementation of \ctpo scales as ${\mathcal O}(exp(\max\{\NumV,\log^5(\maxcap)\}))$, and that the redundancy introduced by such codes scales as ${\mathcal O}(\NumV exp(\max\{\NumV,\log^5(\maxcap)\}))$.
}

The problem of polynomial identity testing (PIT)~\cite{AgrS:09} examines the question of deterministically determining whether a polynomial with a succinct but non-standard representation (such as the determinant of a matrix of polynomials) identically equals zero. The deterministic complexity of such problems is a long-standing open problem in theoretical computer science. Given this context, we are unable to provide intuition on whether our codes in Section~\ref{subsec:c3p0} have order-optimal computational complexity -- indeed, answering this question in either direction would represent significant progress in resolving the complexity of PIT problems.


\section{Implementation issues}
{
As noted in~\cite{JagEHM:04}, the complexity of implementation of network codes scales polynomially in the logarithm of the field-size over which operations are performed, or in the case of convolutional network codes, polynomially in the degree of the polynomials used at each node. By this measure, the implementation complexity of the codes in~\cite{HoKMKE:03,JagCJ:03} is poly-logarithmic in network parameters, whereas the implementation complexity of the first three of the four codes in this work is polynomial in network parameters. While this is an exponential blow-up, we note that the resulting codes are still computationally tractable, and further, as noted in Section~\ref{prob_complexity}, such a blow-up is in fact necessary for codes to be universal. 

\begin{table}[ht]
\centering 
\begin{tabular}{|c | c | c | c | c | }
\hline
   Code &  $\NumS$ &  $\NumE$ & Error & Implementation
   \\
   &  known &  known & Probability & Complexity
   \\
\hline 
\cite{HoKMKE:03} & yes & yes&  $\epsilon$ & ${\mathcal O}(polylog(\NumE\NumS))$
\\
\hline 
\cite{LangbergSB:05} & yes & yes& 0 &${{\mathcal O}}(polylog(\NumS))$ 
\\
\hline 
\wupp & yes & no & $\epsilon$ & ${\mathcal O}(poly(\NumV\log(\maxcap))$ 
\\
& & & & $+ \log(\Rate\NumS/\epsilon))$ 
\\
\hline 
\supp & no & no & $\epsilon$ &${\mathcal O}(poly(\Rate\NumV\log(\maxcap))$
\\
& & & & $+ \log(1/\epsilon))$ 
\\
\hline 
\rtdt & no & no & 0 & ${\mathcal O}(\max\{\NumV,\log^3(\maxcap)\})$
\\
\hline 
\ctpo & no & no & 0 & ${\mathcal O}({exp(\max\{\NumV, \log^5(\maxcap)\})})$
\\
\hline 
\end{tabular}
\label{table_comparison}
\caption{{A summary of the properties of our constructions, and comparison between them and prior non-universal algorithms. }}
\end{table}

While the schemes in this work have been presented in the context of convolutional network coding operations at each node, they also go through for other infinite fields such as ${\mathbb Q}$ -- the only requirement is that the field be unbounded in size, and that an infinite subset of it have a succinct representation.

Also, despite presenting all messages at the source and each link as bit-streams of possibly unbounded length, the schemes described in prior sections can also be implemented by packetization, by chopping up the bit-streams into packets of a standard size $\Bl$.}

In our codes the header of each packet contains low-rate control information used by each node to decide on its coding operations. However, by design, the size of this header changes as information flows down the network -- the rate of change depends on the network topology, and hence is unpredictable in advance. One challenge in the implementation of our codes is thus to ensure that the intermediate nodes are able to distinguish between header information and payload information. One standard trick for such scenarios is used in~Theorem $14.2.3$ of~\cite{CoverT:06} -- each bit of the header is doubled, and the final such double-bit is followed by a $01$ to signify the end of the header. Since the length of the header is asymptotically negligible in the packet-size, the communication cost of this bit-doubling is still asymptotically negligible.

\section{Discussion}
{In this work we provide the first rate-optimal network code designs that have guaranteed decodability performance, and yet are independent of all network parameters. While requiring such universality makes us pay a price in the computational complexity and redundancy, (all but one of) our codes are computationally efficient to implement. The analytical tools we derive may well be of independent interest.}

\bibliographystyle{IEEEtran}
\bibliography{universal}

\appendix \label{sec:app}
\subsection{Proof of Lemma~\ref{lem:SZ_Extend1}} \label{sec:app1}
We proceed by mathematical induction. In the base case when $\dummy=1$, Lemma~\ref{lem:SZ_Extend1} is equivalent to the Schwartz-Zippel lemma in one variable. 

As the inductive hypothesis, suppose that Lemma~\ref{lem:SZ_Extend1} is true for $\dummy -1$ variables in the polynomial $P(.)$. 

Now consider the case when the polynomial $P(x_1,x_2,\dots,x_\dummy)$ has $\dummy$ variables.
The polynomial can be rewritten so that
$$
P(x_1,x_2,\dots,x_\dummy)= x_\dummy^{d_\dummy}P_1(x_{1},x_{2},\ldots,x_{\dummy -1})
+R_1(x_{1},x_{2},\ldots, x_{\dummy})
$$
\noindent for some polynomials $P_1(.)$ and $R_1(.)$ over the appropriate variables. 

The probability that $P(.)$ equals zero can be bounded from above by
\begin{eqnarray}
\Pr[P(.)=0] & =& \Pr[P(.)=0,P_1(.)=0]+\Pr[P(.)=0,P_1(.)\neq 0]\nonumber\\
&=&\Pr[P_1(.)=0]\Pr[P(.)=0|P_1(.)=0]\nonumber\\
&&+\Pr[P_1(.)\neq 0]\Pr[P(.)=0|P_1(.)\neq 0]\nonumber\\
&\leq&\Pr[P_1(.)=0]+\Pr[P(.)=0|P_1(.)\neq 0] \label{eq:prp}
\end{eqnarray}

But by the inductive hypothesis
\begin{equation}
\Pr [P_1(.) =0] \leq \sum_{i = 1}^{\dummy-1} \frac{\dg_i}{|S_i|}. 
\label{eq:prp1}
\end{equation}

Also, by the Principle of Deferred Decisions~\cite{MitzUp:05} the probability $\Pr(P(.)=0)$ is unaffected if the value of $x_{\dummy}$ is chosen after the values of all the other variables have been fixed. In this case, if $P_1(.)\neq 0$, then $P(.)$ is a polynomial of degree $\dg_\dummy$ over $x_\dummy$. By the Schwartz-Zippel lemma
\begin{equation}
\Pr[P(.)=0 | P_1(.) \neq 0]\leq\frac{\dg_\dummy}{|S_\dummy|}.
\label{eq:prpp1}
\end{equation}

Substituting (\ref{eq:prp1}) and (\ref{eq:prpp1}) into (\ref{eq:prp}) gives the required result.
\hfill $\Box$


\end{document}